\documentclass[12pt,a4paper]{article}
\usepackage{a4wide}
\usepackage{latexsym}
\usepackage{epsf}
\usepackage{amssymb}
\begin{document}
\def\be{\begin{equation}}
\def\ee{\end{equation}}
\def\bea{\begin{eqnarray}}
\def\eea{\end{eqnarray}}

\def\pd{\partial}
\def\a{\alpha}
\def\b{\beta}
\def\g{\gamma}
\def\d{\delta}
\def\m{\mu}
\def\n{\nu}
\def\t{\tau}
\def\l{\lambda}

\def\s{\sigma}
\def\e{\epsilon}
\def\scri{\mathcal{J}}
\def\cM{\mathcal{M}}
\def\tcM{\tilde{\mathcal{M}}}
\def\RR{\mathbb{R}}
%%%%%%%%%%%%%%%%%%%%%%%%%%%%%%%%%%%%%%%%%%%%%%%%%%%%%%%%%%%%%%%%%%%%%%

\hyphenation{re-pa-ra-me-tri-za-tion}
\hyphenation{trans-for-ma-tions}

%%%%%%%%%%%%%%%%%%%%%%%%%%%%%%%%%%%%%%%%%%%%%%%%%%%%%%%%%%%%%%%%%%%%%%

\begin{flushright}
IFT-UAM/CSIC-00-14\\
hep-th/0003205\\
\end{flushright}

\vspace{1cm}

\begin{center}

{\bf ON THE STRING DESCRIPTION OF CONFINEMENT }

\vspace{.5cm}
 
{\bf Enrique \'Alvarez}
\footnote{E-mail: {\tt enrique.alvarez@uam.es}}
and {\bf C\'esar G\'omez}
\footnote{E-mail: {\tt cesar.gomez@uam.es}}
\vspace{.3cm}

\vskip 0.4cm

{\it
 Instituto de F\'{\i}sica Te\'orica, C-XVI,
\footnote{Unidad de Investigaci\'on Asociada
  al Centro de F\'{\i}sica Miguel Catal\'an (C.S.I.C.)}
and  Departamento de F\'{\i}sica Te\'orica, C-XI,\\
  Universidad Aut\'onoma de Madrid 
  E-28049-Madrid, Spain }

\vskip 0.2cm

\vskip 1cm

%%%%%%%%%%%%%%%%%%%%%%%%%%%%%%%%%%%%%%%%%%%%%%%%%%%%%%%%%%%%%%%%%%%%%%

{\bf Abstract}
A non supersymmetric string background ,directly derived from
the string soft dilaton theorem, is used to compute, in the semiclassical
approximation, the expectation value of Wilson loops in static gauge. 
The resulting potential shares common features with the one
obtained through Schwarzschild-anti de Sitter spacetime metrics 
 In particular a linear confining potential appears naturally.

\end{center}
%%%%%%%%%%%%%%%%%%%%%%%%%%%%%%%%%%%%%%%%%%%%%%%%%%%%%%%%%%%%%%%%%%%%%%

\begin{quote}

\end{quote}

%%%%%%%%%%%%%%%%%%%%%%%%%%%%%%%%%%%%%%%%%%%%%%%%%%%%%%%%%%%%%%%%%%%%%%

\newpage
%%%%%%%%%%%%%%%%%%%%%%%%%%%%%%%%%%%%%%%%%%%%%%%%%%%%%%%%%%%%%%%%%%%%%%

\setcounter{page}{1}
\setcounter{footnote}{1}
%\renewcommand{\theequation}{\thesection.\arabic{equation}}
%\tableofcontents
\newpage

\vspace{1cm}
%%%%%%%%%%%%%%%%%%%%%%%%%%%%%%%%%%%%%%%%%%%%%%%%%%%%%%%%%%%%%%%%%%
\section{Introduction}
%%%%%%%%%%%%%%%%%%%%%%%%%%%%%%%%%%%%%%%%%%%%%%%%%%%%%%%%%%%%
%%%%%%%%%%%%%%%%%%%%%%%%%%%%%%%%%%%%%%%%%%%%%%%%%%%%%%%%%%%%%%%%%%%
The simplest candidate for a string representing 
 pure Yang Mills theory, is a non
critical string with \emph{curved Liouville} \cite{Polyakov1} action:
\be\label{action}
L= (\frac{a(\phi)}{l_{s}})^{2}(\partial x)^{2} + (\partial \phi)^{2} 
+ T(\phi) + \Phi(\phi) R_{2}
\ee
where $x$ stands for the four dimensional space-time coordinates, 
$\phi$ for the Liouville 
field and $T(\phi)$ and $\Phi(\phi)$ for the closed tachyon and dilaton 
backgrounds. The factor $\frac{a(\phi)}{l_{s}}$ would be interpreted as
an effective \emph{running} string tension for the four dimensional
non critical string \cite{Polyakov2}. The scale $l_{s}$ plays then the
r\^ole of 
a \emph{bare} string tension. The space-time metric associated to the preceding
action (\ref{action}) is:
\be\label{metric}
ds^{2} = a(\phi)^{2}d x^{2} + l_{c}^{2} d{\phi}^{2}
\ee
where we have introduced an extra scale $l_{c}$ for dimensional reasons. 
The physical meaning of this scale will become clear as we proceed.
\par
The physical backgrounds $a(\phi)$, $T(\phi)$ and $\Phi(\phi)$ 
should be restricted by 
demanding the vanishing of the two dimensional sigma model beta functions.

\par

The soft dilaton theorem \cite{ademollo} for vanishing dilaton tadpoles
(owing to conformal invariance) reads:
\be\label{dt}
(\sqrt\alpha'\frac{\partial}
{\partial{\sqrt\alpha'}}-\frac{1}{2}(d-2)g\frac{\partial}{\partial{g}})
A(p_{1},p_{2},...p_{n})=0
\ee
This equation (\ref{dt})can be considered, for $d=4$, as a renormalization
group equation \footnote{
In closed string field theory the soft dilaton theorem becomes equivalent
to the invariance of the string field action under space-time
dilatations and changes of the string coupling (see \cite{Hata} and 
\cite{barton})} with:
\be\label{beta}
\sqrt\alpha'\frac{\partial g}{\partial{\sqrt\alpha'}}=\beta(g)= -g
\ee
Using the definition of $g$ in terms of the dilaton field:
\be\label{g}
e^{\Phi} =g
\ee
and interpreting , as discussed above, $\sqrt\alpha'$ as $\frac{l_{s}}
{a(\phi)}$ we get from (\ref{beta}) \footnote{After these identifications
(\ref{dt}) becomes equivalent \cite{ag1} to the holographic 
renormalization group \cite{verlinde}}:
\be\label{dilaton}
\Phi=log(\frac{a(\phi)}{l_{s}})
\ee
We will take (\ref{dilaton}) as the starting point to determine 
the backgrounds in
(\ref{action}). Vanishing of  the sigma model beta functions 
(to first order) leads to  the following solution:
\be\label{metric}
ds^{2} = \phi\, d{\vec{x}}^{2} + l_{c}^{2} d{\phi}^{2}
\ee

\be\label{dil}
\Phi(\phi) = \frac{1}{2} log (\phi)
\ee
where we have fine tunned the closed string tachyon vacuum 
expectation value to compensate the central charge deficiency in the dilaton 
beta function equation.
A first analysis of the stability of this solution was presented in 
\cite{ag2}.

\par

In this letter we will consider the problem of confinement
for the background metric (\ref{metric}), by explicit computation in
the semiclassical approximation of the Wilson loop vacuum expectation value.

%%%%%%%%%%%%%%%%%%%%%%%%%%%%%%%%%%%%%%%%%%%%%%%%%%%%%%%%%%%

\section{Wilson Loop}
%%%%%%%%%%%%%%%%%%%%%%%%%%%%%%%%%%%%%%%%%%%%%%%%%%%%%%%%%%%

An interesting feature of the preceding metric (\ref{metric}) 
is the existence of a 
\emph{naked} singularity at $\phi=0$. From (\ref{g}) and (\ref{metric}) this
corresponds to the weakly coupled regime of the dual gauge theory. The
boundary of space-time (\ref{metric}) is at $\phi=\infty$. 
\par
In order
to compute the Wilson loop we will follow a Nambu-Goto semiclassical
approximation in static gauge \cite{maldacena}\cite{rey}. We then indentify
\bea
t&=&\tau\nonumber\\
x&=&\sigma
\eea
The induced metric on the world sheet will then be, for static configurations,
\be
ds^2 = \phi\, d\tau^2 +(\phi + l_s^2\phi'^2)d\sigma^2
\ee
and the action reads:
\be
S=\frac{T}{{l_{s}}^{2}}\int_{0}^{1} d\sigma 
\sqrt{\phi(\phi + l_{c}^{2}\phi'^{2})} 
\ee
We will consider U-shape string configurations with Dirichlet
boundary conditions at the codimension one hyersurface 
$\phi = \Lambda$. Denoting by $\phi_{0}$
the tip of the U-shape string we get:
\be\label{energy}
\frac{\phi^{2 }}{\sqrt{\phi(\phi + l_{c}^{2}\phi'^{2})}}= \phi_{0}
\ee
From (\ref{energy}) we easily get for a loop of size $L$ the relation:
\be
L=2l_{c}\phi_{0}^{\frac{1}{2}} \int_{1}^{\sqrt{\frac{\Lambda}{\phi_{0}}}}
\frac{d\xi}{\sqrt{\xi^{4}-1}}
\ee
The action 
is then given by:
\be\label{integral}
S = \frac{2Tl_{c}\phi_{0}^{\frac{3}{2}}}{l_{s}^{2}}
\int_{1}^{\sqrt{\frac{\Lambda}{\phi_{0}}}}
\frac{\xi^{4} d\xi}{\sqrt{\xi^{4}-1}}
\ee
Please notice that the integral (\ref{integral}) is divergent in the limit 
$\Lambda = \infty$. In terms of elliptic functions we get:
\be\label{L}
L=\frac{2 l_{c} \phi_{0}^{\frac{1}{2}}}{\sqrt{2}} F( cos^{-1} \sqrt
{\frac{\phi_{0}}{\Lambda}}, \frac{1}{\sqrt{2}})
\ee
\be
S= \frac{2Tl_{c}\phi_{0}^{\frac{3}{2}}}{l_{s}^{2}} (
\frac{1}{3\sqrt{2}} F( arcos \sqrt
{\frac{\phi_{0}}{\Lambda}}, \frac{1}{\sqrt{2}}) + 
\frac{1}{3} \sqrt{\frac{\Lambda}{\phi_{0}}} \sqrt{(\frac{\Lambda^{2}}
{\phi_{0}^{2}} -1)})
\ee
From the first equation (\ref{L}) we can read the relationship
 between $\frac{L}{\sqrt{\Lambda}}$
and $ \sqrt{\frac{\phi_{0}}{\Lambda}}$ as plotted in Fig 1.

%%%%%%%%%%%%%%%%%%%%%%%%%%%%%%%%%%%%%%%%%%%%%%%%%%%%%%
\begin{figure}[!ht] 
\begin{center} 
  \leavevmode \epsfxsize= 9cm \epsffile{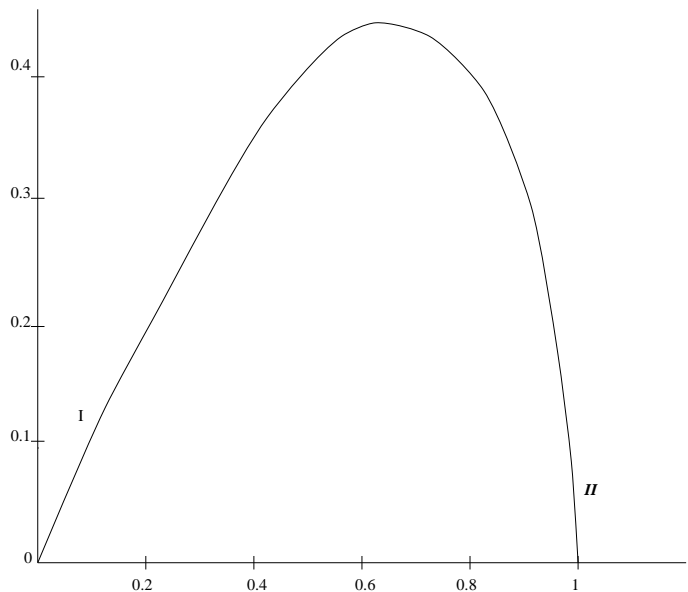}
\caption{ $L/(\sqrt{\Lambda})l_c $ versus $\sqrt{\frac{\phi_0}{\Lambda}}$ }
\label{fig:two} 
\end{center} 
\end{figure}

There are several interesting features. First of all the existence 
of a maximun indicates that the size of the loop $L$ should be necessarily 
smaller that $\sqrt{\Lambda}$ in $l_{c}$ units. Secondly for a given $L$
we get two different U-shape string configurations (see Fig 2).
%%%%%%%%%%%%%%%%%%%%%%%%%%%%%%%%%%%%%%%%%%%%%%%%%%%%%%

\begin{figure}[!ht] 
\begin{center} 
  \leavevmode \epsfxsize= 9cm \epsffile{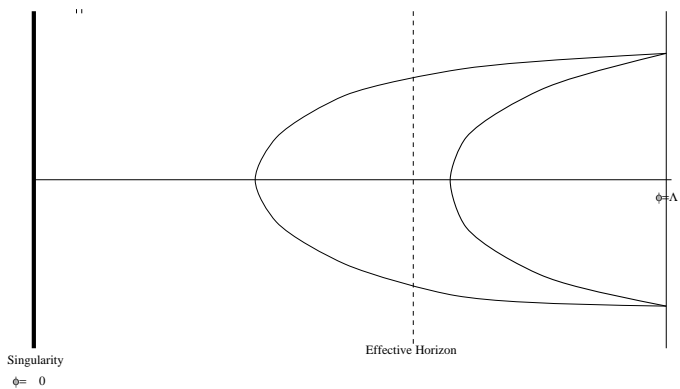}
\caption{ The two U-shaped string configurations corresponding to the same
value of $L$}
\label{fig:two} 
\end{center} 
\end{figure}

This phenomenon
is similar to the one found in \cite{Rey} for Schwarzschild-anti de Sitter
(S-AdS)
space-time ( see Fig 3 for a qualitative comparison ). In fact in \cite{Rey}
a maximum was also obtained (in the context of the S-AdS 
space-time) as well as two possible U-string
configurations for each loop size $L$. However it is important
to stress that in our case this phenomenon depends on having the cutoff
$\Lambda$. In fact the \emph{na\"ive} limit $\Lambda=\infty$ would
 produce the relationship:
\be
L=\frac{2 l_{c} \phi_{0}^{\frac{1}{2}}}{\sqrt{2}} F( \frac{\pi}{2},
\frac{1}{\sqrt{2}})
\ee
%%%%%%%%%%%%%%%%%%%%%%%%%%%%%%%%%%%%%%%%%%%%%%%%%%%%%%%
\begin{figure}[!ht] 
\begin{center} 
  \leavevmode \epsfxsize= 9cm \epsffile{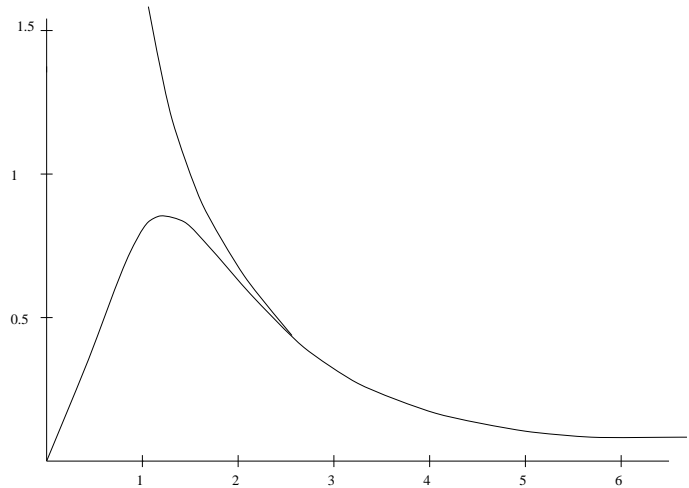}
\caption{ Relation between $L$ and the location of the tip of the U-String configuration for
SadS space-time \cite{Rey}. Also included the curve for AdS space-time.}
\label{fig:two} 
\end{center} 
\end{figure}

Relying upon the similarity with the Schwarzschild-anti de Sitter example we
can think of the vertical dashed line in Fig 3 as a sort of
\emph{effective horizon} covering the singularity at $\phi=0$. Please notice
that the limit $\sqrt{\frac{\phi_{0}}{\Lambda}} =0$ corresponds to pushing
 this
horizon on top of the singularity itself. Hence it looks as if  the system
sort of regularizes  the singularity through a non trivial dependence of
$\phi_{0}$ on $\Lambda$.

\par

Coming back to Fig 3 let us briefly recall the physical meaning of the two
branches. Region $I$ corresponds (see \cite{Rey}) to a confining behavior
for the quark potential while  region $II$ describe the Coulomb phase typical
of $AdS_{5}$. Qualitatively in the deep region $II$, Wilson loops
are defined in terms of electric flux tubes that enter only in the 
asymptotically AdS region reproducing the result of $N=4$ super Yang- Mills.

\par

In our case we can also differenciate between regions $I$ and $II$ of Fig 2.
The static potential in region $I$ corresponding to $\phi_{0} \ll \Lambda$ is
given by:
\be\label{potential}
V = \frac{ L^{3}}{24 K^2 l_{s}^2 l_c^2} + 
\frac{4}{3}\frac{l_c}{l_s^2}\Lambda^{3/2}
\ee
(where $K\equiv K(k=1/\sqrt{2})$ is the complete elliptic integral 
of the first kind), 
This physically means an \emph{overconfining} $L^{3}$ potential between static
probes. It is important 
to stress that the divergent part in (\ref{potential}) cannot be directly
interpreted as a \emph{mass renormalization}.

\par

In the region $II$ for $\phi_{0}$ close to $\Lambda$ and $L \ll \Lambda $ ,in
$l_{c}$ units \footnote{ Notice that here $l_{c}$ is playing the role
of $1/\Lambda_{QCD}$.}, we
get:
\be\label{conf}
V = \frac{L\Lambda}{l_s^2}(1+\frac{\pi\sqrt{2}}{2 K})
\ee
i.e a linear confining behavior. Using equations (\ref{g}) and (\ref{dil})
the leading term of the potential
can be rewritten in terms of the \emph{runnig} effective coupling $g$ as:
\be
V = \frac{Lg^{2}}{l_s^2} (1+\frac{\pi\sqrt{2}}{2 K})
\ee
i.e  a string tension of the order $\frac{g^{2}}{l_s^2}$.

\par
Given a physical value of $L$, it is possible, for each $\Lambda$, to
determine which of the two allowed values of $\phi_0$ (corresponding to
region $I$ or region $II$, respectively), gives smaller potential energy.
In the region in which the approximations are valid it can be written
\be
L = \epsilon l_c \Lambda^{1/2}
\ee
in such a way that
\be
\frac{V_{II}}{V_{I}}=\frac{\epsilon}{\frac{\epsilon^3}{24 K^2}+\frac{4}{3}}
\ee
%%%%%%%%%%%%%%%%%%%%%%%%%%%%%%%%%%%%%%%%%%%%%%%%%%%%%%%%%%%
\section{Comments}
%%%%%%%%%%%%%%%%%%%%%%%%%%%%%%%%%%%%%%%%%%%%%%%%%%%%%%%%%%%

It is natural to expect a renormalization group equation for the Wilson loop
of the type:
\be\label{rge}
(\lambda \frac{\partial}{\partial \lambda} + \beta(g) \frac{\partial}
{\partial g}) W(C) =0
\ee
for dilatations $x \to \lambda x$. The Wilson loop we get satisfy this equation
for the beta function (\ref{beta}). This is the beta function governing the
string field theory. It would be extremely interesting to unravel the relation
between equations of type (\ref{rge}) and the loop equations.

\par

In summary in this letter we have presented a gravitational framework to
study confinement in non supersymmetric Yang Mills. The gravitational
background is dictated by the closed-open relation in string theory as encoded
in the soft dilaton theorem. The dynamics of Wilson loop
on this background shares
many of the features previously found in descriptions of confinement using
generalizations of AdS/CFT correspondence \cite{witten}\cite{Rey}, \cite{cobi}

\section*{Acknowledgments}
This work has been partially supported by the
European Union TMR programs FMRX-CT96-0012 {\sl Integrability,
  Non-perturbative Effects, and Symmetry in Quantum Field Theory}
and  ERBFMRX-CT96-0090 {\sl 
Beyond the Standard model} as well as
by the Spanish grants AEN96-1655 and AEN96-1664.

               \
\end{document}